\documentclass[debug]{rmaa}

\usepackage{paralist}
\usepackage{amsmath}

\usepackage{psfrag,color}

\usepackage[latin1]{inputenc}

\title{Further properties of the dynamical clock A$+$ indicator in a toy model of pure dynamical friction} 

\author{
  M. Pasquato,\altaffilmark{1,2}}

\altaffiltext{1}{INAF, Osservatorio Astronomico di Padova, vicolo dell'Osservatorio 5, I--35122 Padova, Italy.}
\altaffiltext{2}{INFN, Sezione di Padova, Via Marzolo 8, I--35131 Padova, Italy.}

\shortauthor{Pasquato}
\shorttitle{Analytical A$+$ indicator properties}

\fulladdresses{
\item M. Pasquato, INAF, Osservatorio Astronomico di Padova, vicolo dell'Osservatorio 5, I--35122 Padova, Italy.
}

\listofauthors{M. Pasquato}
\indexauthor{Pasquato, M.}

\abstract{The Alessandrini A$+$ indicator is a quantitative measure of star cluster dynamical evolution based on the mass-segregation of blue straggler stars. A$+$ is defined as the integral of the cumulative distribution of blue straggler stars with the radius measured in logarithmic scale, minus a term related to the reference population used. In a companion paper I introduced a simplified model of dynamical friction and calculated the A$+$ indicator analytically. Here I show further properties of the time evolution of A$+$, focusing in particular on the physical interpretation of its time derivative $d{A^+}/dt$. I find that $d{A^+}/dt$ is the mean of the reciprocal dyamical friction timescale, weighted by the density of blue straggler stars. I show that it is non-negative (as expected based on monotonicity) due to the density of blue-straggler stars being non-negative and that, for a dynamical friction timescale that is non-decreasing with radius, $d{A^+}/dt$ is also non-decreasing with time, making the A$+$ indicator a convex function.}

\resumen{}

\addkeyword{Galaxy: globular clusters: general}
\addkeyword{Methods: analytical}
\addkeyword{Stars: blue stragglers}

\begin{document}
% Typeset article header
\maketitle

\section{Introduction}
\label{sec:intro}
\citet{2016ApJ...833..252A} recently introduced a new parameter, intended as a quantitative measurement of the dynamical age of old star clusters.
This indicator, dubbed A$+$ for short, is based on measuring the radial distribution of massive tracers, such as blue straggler stars \citep[][]{1953AJ.....58...61S}, which are undergoing mass segregation by virtue of being heavier that the average stars in the host cluster.
The A$+$ indicator in particular involves taking the integral of the cumulative distribution of blue straggler stars as a function of log radius, minus the same for a reference population, such as horizontal branch stars. The A$+$ indicator is easier to measure compared to other attempts at building a dynamical clock out of the blue straggler star distribution \citep{2012Natur.492..393F, 2013MNRAS.429.1221H, 2015ApJ...799...44M, 2017MNRAS.471.2537H, 2018ApJ...867..163P, 2019MNRAS.483.1523S} as it does not depend on finding the exact position of the minimum of said distribution. While it was originally introduced in the context of N-body simulations, the A$+$ indicator was quickly shown to be able to capture the dynamical evolutionary stage of observed globular clusters, displaying a strong anticorrelation with their half-mass relaxation time \citep{2016ApJ...833L..29L}.

In a companion paper (Pasquato 2019, referred as P19 for short in the following), I worked out an analytical formula for the (three dimensional) A$+$ indicator in a simplified model of mass segregation under pure dynamical friction, and showed that it evolves monotonically as a function of time. Here I build on that to obtain a few interesting properties of the A$+$ indicator. In particular I show that its derivative is a non-negative function, which is non-decreasing as a function of time if the dynamical friction timescale is non-decreasing with radius, making the A$+$ indicator a convex function of time. I provide a simple formula for the value of the A$+$ indicator derivative at the beginning of dynamical evolution, which directly connects it to the dynamical friction timescale. Finally, I suggest that in the late stages of dynamical evolution the A$+$ indicator derivative should approach a constant, making the A$+$ indicator asympthotically linear.

\section{Calculations}
In the contex of the simplified model of evolution under dynamical friction introduced in P19, the A$+$ indicator is calculated as
 \begin{equation}
\label{aless}
A^{+}(t) = \int_{-\infty}^{+\infty} N(I^{-1}(I(e^s) + t), 0) ds - \int_{-\infty}^{+\infty} N(e^s, 0) ds
\end{equation}
where $N(r,t)$ is the cumulative distribution of blue straggler stars as a function of radius and time.
The function $I(\cdot)$ is defined by P19 as
\begin{equation}
\label{integral}
I(r) = \int \frac{\tau (x) dx}{x}
\end{equation}
where $\tau(r)$ is the dynamical friction timescale as a function of radius.

\subsection{Time derivative of the A$+$ indicator}
The second term in Eq.~\ref{aless} does not depend on time, so its derivative vanishes. Regarding the first term, I redefine for simplicity
\begin{equation}
N_0(r) = N(r,0)
\end{equation}
so
\begin{equation}
\label{alessder}
\frac{dA^{+}}{dt} = \frac{d}{dt} \int_{-\infty}^{+\infty} N_0(I^{-1}(I(e^s) + t)) ds =  \int_{-\infty}^{+\infty} \frac{d}{dt} N_0(I^{-1}(I(e^s) + t)) ds
\end{equation}
which, by using the inverse function derivative rule, becomes
\begin{equation}
\label{alessdermore}
\frac{dA^{+}}{dt}  = \int_{-\infty}^{+\infty} \frac{d N_0}{dr}\left( I^{-1}(I(e^s) + t) \right) \frac{ds}{I^\prime(I^{-1}(I(e^s) + t))} 
\end{equation}
that is
\begin{equation}
\label{alessderevenmore}
\frac{dA^{+}}{dt}  = \int_{-\infty}^{+\infty} \frac{d N_0}{dr}\left( r_0(e^s, t) \right) \frac{r_0(e^s, t)}{\tau(r_0(e^s, t))} ds
\end{equation}
and substituting back $s = \log r$ I get
\begin{equation}
\label{alessdernice}
\frac{dA^{+}}{dt}  = \int_{0}^{+\infty} \frac{d N_0}{dr}\left( r_0(r, t) \right) \frac{r_0(r, t)}{r \tau(r_0(r, t))} dr
\end{equation}
by further substituting, for a given $t$, $r_0 = r_0(r, t)$, I get
\begin{equation}
\label{alessdernicenice}
\frac{dA^{+}}{dt}  = \int_{0}^{+\infty} \frac{d N_0}{dr}(r_0) \frac{r_0}{r(r_0, t) \tau(r_0)} \frac{dr}{dr_0} dr_0 = \int_{0}^{+\infty} \frac{d N_0}{dr}(r_0) \frac{ds}{ds_0} \frac{d r_0}{\tau(r_0)}
\end{equation}
where $s_0 = \log r_0$. Alternatively, I can use the inverse function derivative again to obtain
\begin{equation}
\label{alessdernicenicenice}
\frac{dA^{+}}{dt}  =  \int_{0}^{+\infty} \frac{d N_0}{dr}(r_0) \frac{dr_0}{\tau(r)} 
\end{equation}
%= \int_{0}^{+\infty} \frac{d N_0}{dr}(r_0) \frac{r_0}{r(r_0, t) \tau(r_0)} \frac{r}{\tau(r)} \frac{\tau(r_0)}{r_0} dr_0 

This is a general formula for the derivative of the A$+$ indicator at time $t$. It is always non-negative because the integrand is non-negative as long as the derivative of the initial cumulative distribution function is: this condition is equivalent to requesting that density is always non-negative. In other words, the monotonicity of the A$+$ indicator derives from the condition that the number density of blue straggler stars is everywhere non-negative.
Note how in Eq.~\ref{alessdernicenicenice} the time dependence is all enclosed within the $\tau(r)$ term (not $\tau(r_0)$!) at the denominator.
If $\tau(r)$ is constant I recover the result of P19 where the A$+$ indicator is linear with time, as its derivative equals $1/\tau(0)$ and its initial value is always zero by construction.

\subsection{Convexity}
Since for any $r_0, t$ it holds that $r(r_0, t) \le r_0$ and so, if $\tau(r)$ is non-decreasing, $\tau(r_0) \ge \tau(r)$ I have
\begin{equation}
\label{alessdernicenicenicest}
\frac{dA^{+}}{dt} =  \int_{0}^{+\infty} \frac{d N_0}{dr}(r_0) \frac{dr_0}{\tau(r)}  \ge  \int_{0}^{+\infty} \frac{d N_0}{dr}(r_0) \frac{dr_0}{\tau(r_0)}
\end{equation}
which is a lower limit to the derivative of the A$+$ indicator.
 
Eq.~\ref{alessdernice} can be used to calculate the slope of the A$+$ indicator at the start of dynamical evolution, e.g. in a simulation, by setting $t=0$.
\begin{equation}
\label{alessdernice0}
\frac{dA^{+}}{dt} \bigg \vert_0 = \int_{0}^{+\infty} \frac{d N_0}{dr}(r) \frac{dr}{\tau(r)}
\end{equation}
or by further substitution
\begin{equation}
\label{alessdernicer00}
\frac{dA^{+}}{dt} \bigg \vert_0  = \int_{0}^{1} \frac{dN_0}{\tau(r(N_0))} = \int_{0}^{1} \frac{dN_0}{\tau_L(N_0)}
\end{equation}
where $\tau_L(\cdot)$ is the Lagrangian dynamical friction timescale function, i.e. the function that returns the dynamical friction timescale for a given Lagrangian radius, e.g. $\tau_L(0.5)$ is the dynamical friction timescale at the half-number radius (for the initial mass distribution or, equivalently under the assumptions made by P19, the distribution of reference stars). So the derivative of the A$+$ indicator at the beginning of dynamical evolution (e.g. at the start of a simulation) is the mean of the reciprocal of the Lagrangian dynamical friction timescale.

By combining Ineq.~\ref{alessdernicenicenicest} with Eq.~\ref{alessdernice0} and using the fact that both $r$ and $r_0$ are just dummy variables in the respective integrals, I obtain
\begin{equation}
\label{alessderlowerlimit}
\frac{dA^{+}}{dt} \ge  \int_{0}^{+\infty} \frac{d N_0}{dr}(r_0) \frac{dr_0}{\tau(r_0)} = \frac{dA^{+}}{dt} \bigg \vert_0
\end{equation}
so at any time after the start the A$+$ indicator increases at least as fast as the beginning.

Moreover, if the function $\tau(r)$ is non-decreasing, then the $1/\tau(r)$ term inside the integral in Eq.~\ref{alessdernicenicenice} is larger ar smaller radii. As all radii shrink with time, i.e. $r(r_0, t_2) < r(r_0, t_1)$ for every $t_2 > t_1$ and every $r_0$, then the $1/\tau(r)$ term becomes larger with time. Thus the time derivative of the A$+$ indicator increases over time in general, i.e. there is nothing special about the beginning of dynamical evolution.
In other words, the A$+$ indicator is a convex function of time.

This is a quite interesting result in the light of Fig.~5 of \citet{2016ApJ...833..252A} where the (two dimensional) A$+$ indicators calculated for different simulations appear to evolve with increasing slope for increasing times, modulo the numeric error. The physical interpretation for this result is that the evolution becomes faster as the blue straggler stars drop to lower radii, where the dynamical friction timescale is shorter.

\subsection{Limiting behaviour at large times}
If I set $\tau_0 = \tau(0) = \tau_L(0)$ and $\tau_\infty = \lim_{r \to \infty} \tau(r) = \tau_L(1)$ I obtain
\begin{equation}
\label{alessdernicer0minor}
\frac{1}{\tau_\infty} < \frac{dA^{+}}{dt} \bigg \vert_0 < \frac{1}{\tau_0}
\end{equation}
by substitution into Eq.~\ref{alessdernicenicenice}, as $\tau(r)$ is increasing with radius. 

In general, since a constant $\tau(r) = \tau(0)$ holds for the approximately constant-density cores of globular clusters, I expect the final stages of the evolution, when most blue stragglers sunk to the core, to have a value of the A$+$ indicator's derivative that approaches the constant value $1/\tau(0)$. As I have shown, any value attained later in evolution should be larger than previous values of the derivative. This is another point where comparison with Fig.~5 of \citet{2016ApJ...833..252A} is reassuring, as all the slopes in that figure are quite smaller than one, even though their values of the A$+$ indicator are referred to the half-mass relaxation timescale, which is longer than the central dynamical friction timescale. Further caveats for a comparison are that:
\begin{itemize}
\item at some point during the evolution of a cluster, my simplified picture of dynamical friction will no longer represent reality as diffusion effects will start to dominate, right in the central regions, so the limiting value of ${dA^{+}}/{dt} = 1/\tau(0)$ is not necessarily reached.
\item blue stragglers on very eccentric orbits already depart from my simplified picture, even in the beginning of their evolution
\item \citet{2016ApJ...833..252A} uses a projected, two-dimensional version of the A$+$ indicator
\item the reference stars (horizontal branch stars) they use are also evolving over time, as they are likely heavier than the average star in the simulations they run
\end{itemize}
In subsequent papers I will address the third point, obtaining a better basis for comparison with simulations. The fourth point is instead addressed in the following.

\subsection{Moving reference stars}
Until now I assumed that the reference stars for the A$+$ indicator are fixed, i.e. they keep the initial cumulative distribution over time. If I relax this assumption, while still assuming that the initial cumulative distribution of blue straggler and reference stars coincides, Eq.~\ref{aless} becomes
\begin{equation}
\label{alessmoving}
A^{+}(t) = \int_{-\infty}^{+\infty} N(I^{-1}(I(e^s) + t), 0) ds - \int_{-\infty}^{+\infty} N(\hat{I}^{-1}(\hat{I}(e^s) + t), 0) ds
\end{equation}
where
\begin{equation}
\label{integral}
\hat{I}(r) = \int \frac{\hat{\tau} (x) dx}{x}
\end{equation}
and $\hat{\tau}(r)$ is the dynamical friction timescale as a function of radius for the reference particles. If $\hat{\tau}(r) \gg {\tau}(r)$ for every $r$ then we revert back to the original situation where I can neglect the effects of dynamical friction on the reference population, while if $\hat{\tau}(r) = {\tau}(r)$ then the A$+$ indicator will be identically zero over time.
In the following I will assume that 
\begin{equation}
\label{bastatelefonateperfavore}
\hat{\tau}(r) = \frac{m_{BSS}}{m_{HB}} {\tau}(r)
\end{equation}
where $m_{BSS}$ is the mass of a blue straggler star and $m_{HB}$ the mass of a reference star.
This is in line with e.g. \citet{2004ApJ...605L..29M}, which assumes that dynamical friction scales inversely with mass.
With this in mind it is easy to see that the time derivative of A$+$ calculated in Eq.~\ref{alessdernicenicenice} becomes
\begin{equation}
\label{alessdernicenicenicemultipop}
\frac{dA^{+}}{dt}  =  \int_{0}^{+\infty} \frac{d N_0}{dr}(r_0) \frac{dr_0}{\tau(r)} - \int_{0}^{+\infty} \frac{d N_0}{dr}(r_0) \frac{dr_0}{\hat{\tau}(r)}
\end{equation}
which, by substituting in Eq.~\ref{bastatelefonateperfavore}, leads to the final result
\begin{equation}
\label{alessdernicenicenicemultipop}
\frac{dA^{+}}{dt}  =  (1 - \frac{m_{HB}}{m_{BSS}}) \int_{0}^{+\infty} \frac{d N_0}{dr}(r_0) \frac{dr_0}{\tau(r)}
\end{equation}
which shows that the slope of the time evolution of A$+$ is merely rescaled by a constant factor. Since both the A$+$ indicator referred to a static population and the one referred to a population that also suffers dynamical friction are $0$ at the beginning of evolution, then A$+$ itself is merely rescaled by a constant factor. Thus all the results I obtained above still hold.

\section{Conclusions}
Working within the same pure dynamical friction picture I introduced in the companion paper P19, I have calculated the time derivative of the \citet{2016ApJ...833..252A} A$+$ indicator and found that it is not decreasing in time if the dynamical friction timescale $\tau(r)$ is non-decreasing with radius. Thus the A$+$ indicator is a convex function of time. P19 results show that in a constant dynamical friction timescale scenario the A$+$ indicator is linear with time, so convexity implies that over time the A$+$ indicator evolves with increasing slope becoming asympthotically linear, as the late stages of evolution take place in the uniform density cores of globular clusters, where the dynamical friction timescale becomes constant.
Finally, I have shown that the choice of a reference population that is also affected by mass-segregation (as is likely the case in an observational setting) merely leads to a constant rescaling of the A$+$ indicator.

\section*{Acknowledgments}
This project has received funding from the European Union's Horizon $2020$
research and innovation programme under the Marie Sk\l{}odowska-Curie grant agreement No. $664931$.
I wish to thank Dr. Pierfrancesco di Cintio, Dr. Paolo Miocchi, Dr. Alessandro Cobbe, and Dr. Stefano Pugnetti for helpful discussion on this subject.

\bibliography{ms2}

\end{document}